\documentclass[%
 reprint,
 amsmath,
 amssymb,
 aps,
 physrev,
 nobibnotes,
 nofootinbib,
 superscriptaddress
]{revtex4-2}

\usepackage{aas_macros}
\usepackage{graphicx}
\usepackage{bm}
\usepackage{multirow}
\usepackage[dvipsnames]{xcolor}
\usepackage[
  setpagesize=false,%
  bookmarks=true,%
  bookmarksdepth=tocdepth,%
  bookmarksnumbered=true,%
  colorlinks=true,%
  linkcolor=blue,
  citecolor=blue,
  urlcolor=blue,
  anchorcolor=magenta,
  linktoc=all
]{hyperref}

\usepackage{ulem}

\newcommand{\nvi}{\nu\mathrm{VI}}
\newcommand{\noct}{\nu\mathrm{OCT}}

\begin{document}

\title{Revisiting constraints on primordial vector modes and implications for sourced magnetic fields and observed $EB$ power spectrum}

\author{Kaito Yura}
 \email{yura.kaito.p8@s.mail.nagoya-u.ac.jp}
\affiliation{Department of Physics, Nagoya University, Furo-cho Chikusa-ku, Nagoya 464-8602, Japan}

\author{Shohei Saga}
\affiliation{Institute for Advanced Research, Nagoya University Furocho, Chikusa-ku, Nagoya, 464-8602, Japan}
\affiliation{Kobayashi Maskawa Institute, Nagoya University, Nagoya, Aichi 464-8602, Japan}

\author{Shuichiro~Yokoyama}
\affiliation{Kobayashi Maskawa Institute, Nagoya University, Nagoya, Aichi 464-8602, Japan}
\affiliation{Department of Physics, Nagoya University, Nagoya, Aichi 464-8602, Japan}
\affiliation{Kavli Institute for the Physics and Mathematics of the Universe (WPI), The University of Tokyo, Kashiwa, Chiba 277-8583, Japan}

\author{Kiyotomo Ichiki}
\affiliation{Department of Physics, Nagoya University, Furo-cho Chikusa-ku, Nagoya 464-8602, Japan}
\affiliation{Kobayashi Maskawa Institute, Nagoya University, Nagoya, Aichi 464-8602, Japan}

\date{\today}

\begin{abstract}
We revisit regular primordial vector modes sustained by the anisotropic stress of free-streaming neutrinos. 
We consider two classes of neutrino-sector initial conditions, the neutrino velocity isocurvature mode ($\nvi$) and the neutrino octupole mode ($\noct$). 
We update their observational constraints using current cosmological data, and examine the impact of including the BICEP/Keck 2018 $B$-mode polarization data.
From an MCMC analysis, we obtain the 95\% C.L. upper bounds on the vector-to-scalar ratio as $r_\mathrm{v}<1.55\times10^{-4}$ and $r_\mathrm{v}<1.04\times10^{-2}$ for the $\nvi$ and $\noct$ modes at the vector pivot scale $k_{0} = 0.01\,{\rm Mpc}^{-1}$, respectively. 
We then study two consequences of these bounds. 
First, we estimate the magnetic fields inevitably generated in the pre-recombination plasma associated with the vector modes. 
We find that the magnetic-field amplitude at recombination with a coherent length of $1~{\rm Mpc}$ is bounded by $B\sim\mathcal{O}(10^{-23})\,{\rm G}$ and $B\sim\mathcal{O}(10^{-21})\,{\rm G}$ for the $\nvi$ and $\noct$ modes, respetively, which is too small to provide the seed of magnetic fields observed today.
Second, assuming the helical vector mode, we compute the induced CMB $EB$ spectrum. 
We show that even a fully helical primordial vector mode cannot reproduce the currently observed $EB$ signal while remaining consistent with parity-even CMB constraints.

\end{abstract}

\maketitle

\section{Introduction}
\label{sec:Introduction}

Although vector perturbations are usually neglected in standard cosmology because they decay with an expanding universe, cosmological models with regular primordial vector modes have been proposed through the non-vanishing anisotropic stress. Such modes constitute a well-defined extension of primordial initial conditions and can leave characteristic imprints in cosmic microwave background (CMB) temperature and polarization anisotropies.
They may be sourced, for example, by primordial magnetic fields through their anisotropic stress, or more generally in scenarios that involve additional vector degrees of freedom and nontrivial gauge field dynamics in the early universe~(see e.g., Ref.~\cite{2013A&ARv..21...62D} for reviews).
Perturbations in the neutrino sector can also play a crucial role. 
Once neutrinos begin to free-stream, their anisotropic stress permits regular primordial vector solutions that can survive and leave observable imprints~\cite{1992ApJ...392..385R,2004PhRvD..70d3518L}. 
In this paper, we focus on those vector modes sourced by neutrino sector initial conditions, which are well motivated as a probe of non-standard sources of anisotropic stress and the associated physics in the early universe.

The most direct observational signatures of primordial vector modes are expected in the CMB temperature and polarization anisotropies. To date, the vector modes have been indeed constrained by the CMB temperature and polarization data~\cite{2012PhRvD..85d3009I,2014JCAP...10..004S}.
Primordial vector modes can also leave imprints on late-time large-scale structure, which have been constrained by several authors~\cite{2013MNRAS.432.2331S,2016JCAP...07..037D,2024PhRvD.109d3520S,2023PhRvD.108l3528C,2024PhRvD.109f3541P}.

Over the past decade, CMB polarization measurements have been significantly improved.
In general, primordial vector modes source not only temperature and $E$-mode polarization, but also $B$-mode polarization~\cite{2004PhRvD..70d3518L}.
Importantly, the behavior of the $B$-mode power spectrum sourced by vector modes differs from that sourced by primordial tensor modes, making polarization data particularly powerful to resolve such degeneracies of the vector modes with tensor modes.
The early analyses~\cite{2012PhRvD..85d3009I} in the \textit{WMAP} era relied on $TT/TE/EE$ and suffered from degeneracies with tensor contributions.
Later work~\cite{2014JCAP...10..004S} showed that including $B$-mode data can significantly improve constraints on primordial vector modes, although the data used in that analysis were later shown to be dominated by galactic dust emission~\cite{2015PhRvL.114j1301B}.
With the further progress of the polarization measurements, we aim to update the constraints on the primordial vector mode using the latest CMB data and reassess the impact of current $B$-mode polarization data, in particular BICEP/Keck 2018 (BK18), on the tightness of the constraints~\cite{2021PhRvL.127o1301A}.

We make use of public inference frameworks for Markov Chain Monte Carlo (MCMC) analysis that incorporate likelihoods of current cosmological observations.
The regular primordial vector mode can be realized with different neutrino-sector initial conditions, depending on the choice of the non-vanishing multipoles of the distribution function for free-streaming neutrinos at the initial time.
While earlier analyses~\cite{2012PhRvD..85d3009I,2014JCAP...10..004S} treated primordial vector modes that have the regular mode induced by the neutrino velocity isocurvature ($\nvi$) initial condition, the recent work~\cite{2025JCAP...10..112K} thoroughly investigated the possible modes induced by the neutrino octupole ($\noct$) initial condition and also the sourced vector mode, and they reported the best-fitting parameter values related to the primordial vector mode obtained with the \texttt{BOBYQA} algorithm~\cite{2018arXiv180400154C,2018arXiv181211343C}.
We will perform an MCMC analysis and present constraints at confidence levels for both the $\nvi$ and the $\noct$ modes as a follow-up to the previous work~\cite{2012PhRvD..85d3009I,2014JCAP...10..004S}.

A further motivation for studying primordial vector modes is that they make concrete secondary predictions: they inevitably generate magnetic fields in the pre-recombination plasma through the small baryon-photon relative velocity induced by imperfect Thomson coupling~\cite{1970MNRAS.147..279H,2005PhRvL..95l1301T,2006Sci...311..827I,2012PhRvD..85d3009I,2015PhRvD..91l3510S}.
This is particularly intriguing because magnetic fields are observed on a wide range of astrophysical scales, from galaxies to galaxy clusters, and may also exist in cosmic voids inferred from TeV-blazar observations~\cite{2010Sci...328...73N,2011A&A...529A.144T,2012ApJ...747L..14V,2011MNRAS.414.3566T,2012ApJ...744L...7T,2013ApJ...771L..42T,Yang_2015,2017ApJ...847...39V}.
However, the origin of these magnetic fields remains unsettled.
Although magnetohydrodynamic dynamos can plausibly amplify magnetic fields up to the $\mu$G level once a seed field is present~\cite{2002RvMP...74..775W}, the origin of such a seed is not explained by the dynamo mechanism itself~\cite{2008RPPh...71d6901K,2019Galax...7...47S}.
Primordial vector modes are therefore of particular interest as a possible early-universe source of such seed fields.
Observational upper limits on the primordial vector-mode amplitude can be directly translated into upper bounds on the amplitude of the generated magnetic fields, providing information on a possible early-Universe contribution to cosmic magnetism.
In this paper, we update the bounds on the primordial vector-mode amplitude using the latest cosmological data.

An additional extension is to consider helical primordial vector modes in a phenomenological way.
If the primordial vector perturbations possess a net helicity, they break parity symmetry and induce non-vanishing parity-odd correlations in the CMB polarization.
Parity-odd CMB spectra from helical sources have been widely discussed, e.g.~in the context of helical primordial magnetic fields~\cite{2004PhRvD..69f3006C,2005PhRvD..71j3006K,2016A&A...594A..19P}.
In particular, helical vector modes can source a non-vanishing $EB$ (and $TB$) cross-correlation, which is identically zero in parity-symmetric universe~\cite{2012PhRvD..85h3004K,2014PhRvD..90h3004K}.
This provides a further observational test in light of recent discussions of a reported $EB$ signal (often interpreted as isotropic cosmic birefringence), although the physical origin is under debate~\cite{2020PhRvL.125v1301M,2022A&A...662A..10E,2022PhRvD.106f3503E,2025arXiv250913654D}.
A variety of explanations based on other parity-violating mechanisms have been explored, including scenarios with primordial chiral gravitational waves~\cite{2022PhRvD.106j3529F} and attempts to account for the signal without introducing new particles beyond the standard model~\cite{2024JHEP...01..057N}; these studies highlight both the interest in and the challenges of explaining the observed parity-odd signal using known physics alone.
In this paper, we explicitly compute the $EB$ power spectrum induced by helical primordial vector modes within our framework and compare it with current CMB polarization measurements.

This paper is organized as follows.
In Sec.~\ref{sec:Eq_vector}, we summarize the basic equations of the primordial vector modes, following Ref.~\cite{1992ApJ...392..385R,2004PhRvD..70d3518L}, and introduce our parameterization of the vector modes, along with the resulting CMB power spectra for both the $\nvi$ and $\noct$ modes.
In Sec.~\ref{sec:Constraints}, we present updated constraints from the latest cosmological observations.
In Sec.~\ref{sec:Implications}, we consider the observational implications from the vector modes.
We translate the constraints on the vector modes into bounds on the magnetic fields generated by the primordial vector modes. 
We also compute the $EB$ cross-correlation expected by a fully helical vector mode and compare it with the current measurement of the $EB$ power spectrum.
Finally, Sec.~\ref{sec:Summary} summarizes our conclusions.
In Appendix~\ref{sec:full_constraints}, we present the contours of the full parameters, including cosmological parameters.
We also perform the analysis, including the tensor mode in Appendix~\ref{sec:vec_with_tensor}.

\section{Equations of primordial vector mode and the power spectrum}
\label{sec:Eq_vector}

\begin{figure*}
    \centering
    \includegraphics[width=0.99\linewidth]{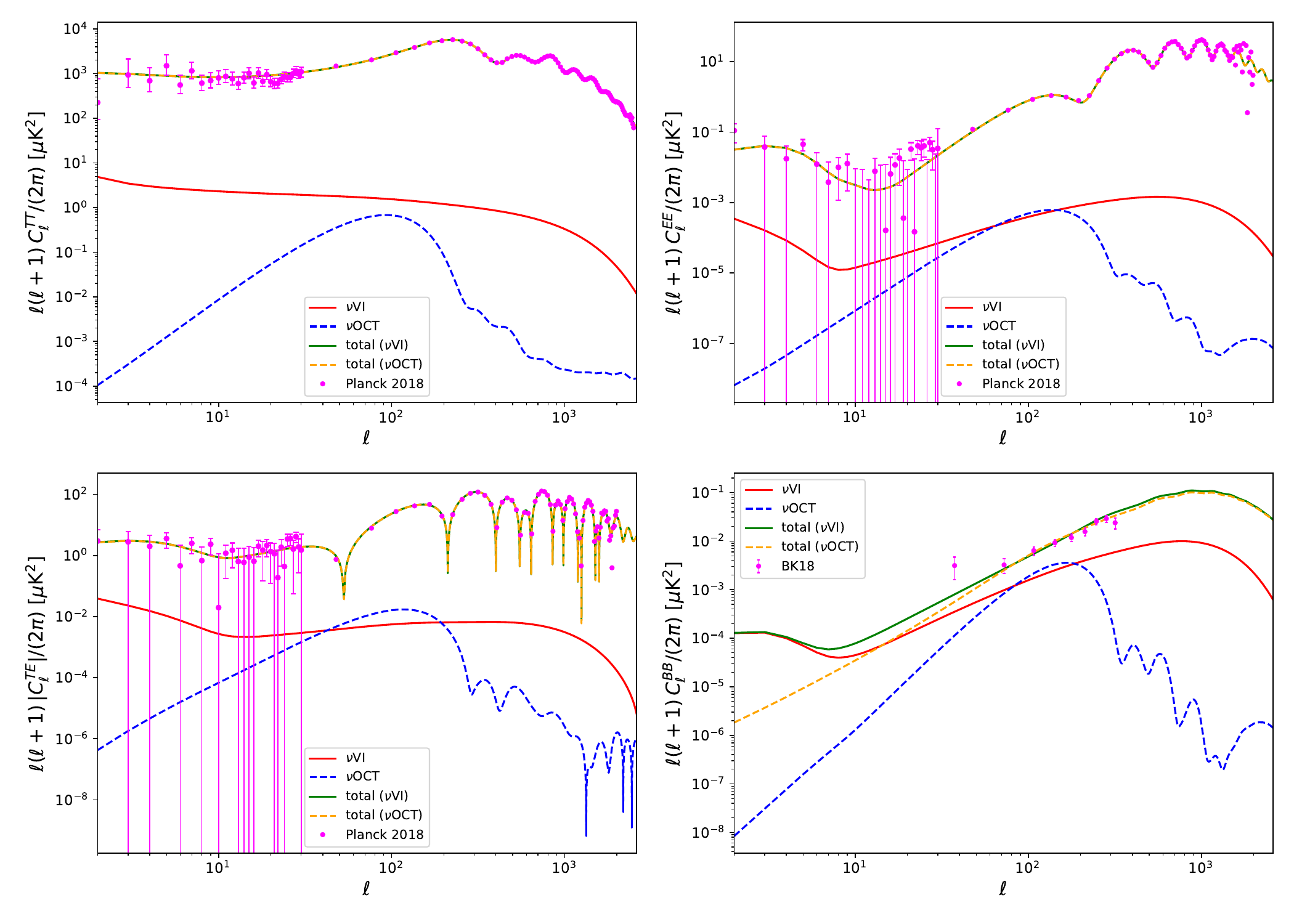}
    \caption{CMB temperature and polarization power spectra ($TT$, $EE$, $BB$) and their cross-correlation ($TE$) sourced by the $\nvi$ and $\noct$ modes.
    For the vector contribution, we choose the best-fit values of $n_{\mathrm{v}}$ and the 95\% C.L.\ upper bounds of $r_{\mathrm{v}}$ as shown later in Sec.~\ref{sec:Constraints}, $(n_\mathrm{v}, r_\mathrm{v})=(0.8,\,1.55\times10^{-4})$ (red solid) and $(n_\mathrm{v}, r_\mathrm{v})=(3.0,\,1.04\times10^{-2})$ (blue dashed) for the $\nvi$ and $\noct$ modes, respectively.
    The total power spectra show the sum of the spectra for the adiabatic scalar curvature perturbations and vector modes, where the standard cosmological parameters are set to be the best-fit values as shown in Appendix.~\ref{sec:full_constraints} for the $\nvi$ and $\noct$ modes, respectively.
    The $BB$ spectrum from scalar perturbations in the lower-right panel is the lensing-induced $B$-mode.
    The data points correspond to \textit{Planck} for $TT$, $EE$, and $TE$, and to BK18 for $BB$.}
    \label{fig:ISO_OCT_Cl}
\end{figure*}

In this section, following Ref.~\cite{2012PhRvD..85d3009I}, we summarize the basic equations governing the evolution of the primordial vector mode. 
We work in synchronous gauge and consider the linear perturbations around a spatially flat Friedmann--Lema\^{i}tre--Robertson--Walker background with the metric
\begin{align}
    ds^2=a(\eta)^2\left[-d\eta^2+(\delta_{ij}+h_{ij})dx^idx^j\right],
\end{align}
where $a(\eta)$ is a scale factor with conformal time $\eta$.

We can decompose metric perturbations into scalar, vector, and tensor modes, which evolve independently at linear order. 
For the vector sector, we expand the metric perturbation in a helicity basis
\begin{align}
    &h_{ij}(\eta,\bm{x})\notag\\
    &=\int\frac{d^3k}{(2\pi)^3}\sum_{\lambda=\pm 1}h^{(\lambda)}(\eta,\bm{k})\left[\hat{k}_{i}\epsilon_{j}^{(\lambda)}(\hat{k})+\hat{k}_{j}\epsilon_{i}^{(\lambda)}(\hat{k})\right]e^{i\bm{k}\cdot\bm{x}},
\end{align}
where the  superscript $\lambda=\pm1$ labels the helicity states.
The quantity $\epsilon_{i}^{(\lambda)}(\hat{k})$ is the divergenceless polarization vector, satisfying the conditions $\hat{k}^i \epsilon_i^{(\lambda)}(\hat{\bm{k}})=0$ and $\epsilon_i^{(\lambda)}(\hat{\bm{k}})\,\epsilon_i^{(\lambda')*}(\hat{\bm{k}})=\delta_{\lambda\lambda'}$.
Unless parity-violating interactions are present, the two helicity states evolve independently. 
In this section, for simplicity, we omit the helicity label, but it will play an important role in Sec.~\ref{sec:EB}.

We define the gauge-invariant vector metric variable $\sigma$ as $\sigma\equiv h^{\prime}/k$, where $\prime$ denotes a derivative with respect to the conformal time $\eta$. 
The linearized Einstein equations in the vector sector are
\begin{align}
    k^2\sigma=-16\pi G a^{2} (\bar{\rho}+\bar{P})v \label{eq: sigma},\\
    \sigma^{\prime}+2\mathcal{H}\sigma=8\pi G a^2 \bar{P} \, \Pi/k
    \label{eq:derivative_sigma},
\end{align}
where the quantities $\mathcal{H}=a'/a$, $\bar{\rho}$, $\bar{P}$, $v$, and $\Pi$ denote the conformal Hubble parameter, the background total energy density, the background total pressure, the vector-mode velocity (vorticity), and the total anisotropic stress, respectively.
In the absence of anisotropic stress, Eq.~\eqref{eq:derivative_sigma} yields the standard decaying solution, $\sigma\propto a^{-2}$.
If, however, free-streaming particles such as neutrinos are present in the early universe, the anisotropic stress can act as a sustained source of vector perturbations and allow the regular primordial vector solution.
In this paper, we consider standard cosmology with free-streaming particles, such as neutrinos. 
Our main focus in this paper is to investigate the cosmological impact of the regular vector mode sourced by the free-streaming neutrinos.

Depending on the initial non-vanishing multipoles of the distribution function of the neutrinos, the regular vector modes associated with the multipoles can arise. In this work, we particularly focus on two initial conditions associated with the lowest and next-lowest nonvanishing neutrino multipoles, commonly referred to as the neutrino velocity isocurvature mode ($\nvi$) and the neutrino octupole mode ($\noct$), respectively.
The $\nvi$ mode is realized by a compensated initial velocity configuration in which photons and neutrinos carry equal and opposite vorticities. 
Through the subsequent evolution, the neutrino anisotropic stress grows and acts as a source for the vector mode. 
By contrast, the $\noct$ mode is characterized by a nonzero initial neutrino octupole perturbation. We can regard this mode as an extension of the usual treatment, in which the initial evolution is discussed by truncating the photon and neutrino multipole hierarchies at the quadrupole level. Introducing an initial octupole perturbation of the neutrino sector provides another way to realize a regular vector solution; although its physical origin remains unclear, it is mathematically well defined.

Here, we define the power spectrum of the vector modes $\sigma$ as
\begin{align}
    \langle\sigma(\bm{k})\sigma^*(\bm{k}')\rangle\equiv(2\pi)^3\delta_{\rm D}(\bm{k}-\bm{k}')\frac{2\pi^2}{k^3}\mathcal{P}_{\sigma}(k)\, ,
    \label{eq:power_sigma}
\end{align}
where the dimensionless primordial power spectrum $\mathcal{P}_{\sigma}$ is parameterized as 
\begin{align}
    \mathcal{P}_{\sigma}(k) = A_{\sigma}\left(\frac{k}{k_0}\right)^{n_\mathrm{v}-1} .
    \label{eq:def_P_sigma} 
\end{align}
The BK18 $B$-mode measurements probe multipoles $40\lesssim \ell \lesssim 400$.
Since a comoving wavenumber $k$ primarily projects onto multipoles $\ell \simeq k\,\chi_\ast$, where $\chi_\ast$ is the comoving distance to last scattering surface, we set $k_{0}=0.01\,{\rm Mpc}^{-1}$ corresponding to $\ell \simeq 140$, which roughly matches to the scales of the BK18 measurements.
When $n_{\mathrm{v}}=1$, the vector mode has a scale-invariant spectrum.
We also define the vector-to-scalar ratio by
\begin{align}
    r_\mathrm{v}\equiv\frac{A_\sigma}{A_\mathrm{s
    }}\, ,
    \label{eq:def_rv}
\end{align}
where the quantity $A_{\mathrm{s}}$ denotes the amplitude of the primordial scalar curvature perturbations.
Note that we choose the pivot scale of the scalar modes to be $k_{\mathrm{s}0}=0.05\mathrm{Mpc}^{-1}$.

Our main analysis includes the contributions of the scalar and vector modes to the CMB temperature anisotropy ($T$) and polarization ($E$, $B$) as our primary interest lies in estimating the maximum amplitude of the vector perturbation in the absence of the tensor mode (see Appendix~\ref{sec:vec_with_tensor} for the analysis with the tensor mode).
Thus, we model the total angular power spectrum of the temperature and polarization fluctuations as
$
C_{\ell}^{XY,\,\mathrm{tot}}=C_{\ell}^{XY,\,\mathrm{S}}+C_{\ell}^{XY,\,\mathrm{V}} ,
$
with $XY \in \{TT, TE, EE, BB\}$. 
The functions $C_{\ell}^{XY,\,\mathrm{S}}$ and $C_{\ell}^{XY,\,\mathrm{V}}$ are the contributions from scalar and vector perturbations, respectively.
To compute $C_{\ell}^{XY,\, \mathrm{V}}$, we modified the \texttt{CAMB} code~\cite{2000ApJ...538..473L}.
Figure~\ref{fig:ISO_OCT_Cl} shows the CMB power spectra sourced by adiabatic scalar curvature perturbations together with the $\nvi$ and $\noct$ modes.
For the contributions from the vector modes, we fix $n_{\mathrm{v}}$ to its best-fit value and set $r_{\mathrm{v}}$ to its 95\% C.L.\ upper bound for each of the $\nvi$ and $\noct$ modes, as will be discussed later in the MCMC analysis.
Note that the scalar contribution to the $BB$ spectrum in Fig.~\ref{fig:ISO_OCT_Cl} arises from gravitational lensing.
The $TT$ and $BB$ power spectra sourced by the $\nvi$ mode are enhanced at low and high multipoles, respectively, whereas those sourced by the $\noct$ mode exhibit a pronounced peak around $\ell\simeq100$. 
Such differences in the spectral shapes can affect the resulting constraints on the vector-mode parameters.

As can be expected from Fig.~\ref{fig:ISO_OCT_Cl}, in particular, since primordial vector modes generate $B$-mode polarization, the inclusion of $B$-mode data should substantially improve the constraints compared with analyses based only on temperature and $E$-mode polarization data.

\section{Updated constraints on the primordial vector mode}
\label{sec:Constraints}

\begin{table*}[t]
  \centering
  \begin{minipage}[t]{0.42\textwidth}
    \centering
    \begin{tabular}{c|c}
      \hline\hline
      Parameter & Prior range \\
      \hline
      $\Omega_b h^2$            & $[0.005:0.1]$ \\
      $\Omega_c h^2$            & $[0.001:0.99]$ \\
      $100\theta_{\mathrm{MC}}$ & $[0.5:10]$ \\
      $\tau$                    & $[0.01:0.8]$ \\
      $\ln(10^{10}A_s)$         & $[1.61:3.91]$ \\
      $n_s$                     & $[0.8:1.2]$ \\
      \hline\hline
    \end{tabular}
  \end{minipage}
  \hfill
  \begin{minipage}[t]{0.52\textwidth}
    \centering
    \begin{tabular}{c|c|c}
      \hline\hline
      Initial condition & Parameter & Prior range \\
      \hline
      \multirow{2}{*}{$\nvi$}
        & $n_{\rm v}$ & $[-3:5]$ \\
        & $r_{\rm v}$ & $[0:0.005]$ \\
      \hline
      \multirow{2}{*}{$\noct$}
        & $n_{\rm v}$ & $[-3:10]$ \\
        & $r_{\rm v}$ & $[0:1]$ \\
      \hline\hline
    \end{tabular}
  \end{minipage}
  \caption{Flat prior ranges adopted in the MCMC analysis. Left: $\Lambda$CDM model parameters. Right: primordial vector-mode parameters for the $\nvi$ and $\noct$ modes.}
  \label{tab:prior_ranges}
\end{table*}

In this section, we present constraints on the vector mode parameters obtained by MCMC analysis.
For the MCMC analysis, we modified the \texttt{cobaya} framework~\cite{2019ascl.soft10019T}.
While Ref.~\cite{2012PhRvD..85d3009I} has used just \textit{WMAP} 7-year data, we utilize the latest CMB datasets, \textit{Planck} and BK18.
In addition, we use publicly available likelihoods for Dark Energy Spectroscopic Instrument (DESI) DR2 BAO and Pantheon Type Ia Supernovae\footnote{We explicitly note the likelihoods used in this analysis: 
$\mathtt{planck\_2018\_lowl.[TT|EE]}$, 
$\mathtt{planck\_2018\_highl\_plik.TTTEEE}$,
$\mathtt{planck\_2018\_lensing.clik}$, 
$\mathtt{bicep\_keck\_2018}$, 
$\mathtt{bao.desi\_dr2}$, and 
$\mathtt{sn.pantheon}$.}.
The prior ranges of the parameters used in the analysis are shown in Table~\ref{tab:prior_ranges}.
The convergence of MCMC chains is assessed using the Gelman-Rubin statistic, and we adopt $R-1 < 0.01$ as the convergence criterion.

\begin{figure}[t]
    \centering
    \includegraphics[width=0.8\linewidth]{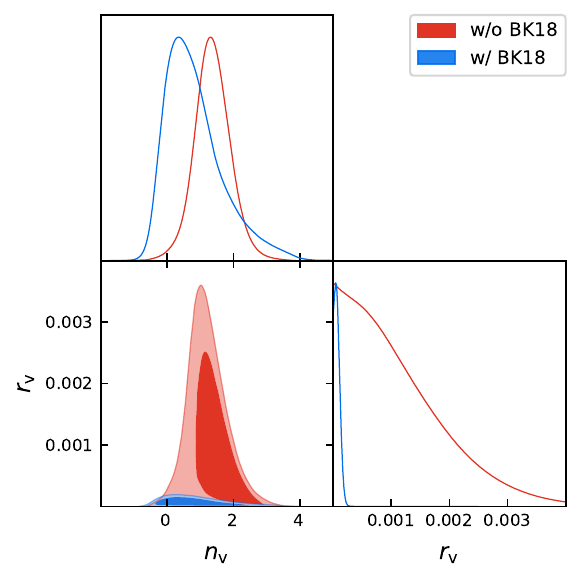}
    \caption{Constraints on the parameters, $n_\mathrm{v}$ and $r_\mathrm{v}$, for the $\nvi$ mode, which were obtained from MCMC analysis.
    The inner and outer contours show the 68\% and 95\% confidence regions, respectively.
    In both cases, we use the \textit{Planck}, DESI and Pantheon data.
    The red and blue contours represent the results without and with BK18 data, respectively.}
    \label{fig:const_iso}
\end{figure}

\begin{figure}[t]
    \centering
    \includegraphics[width=0.8\linewidth]{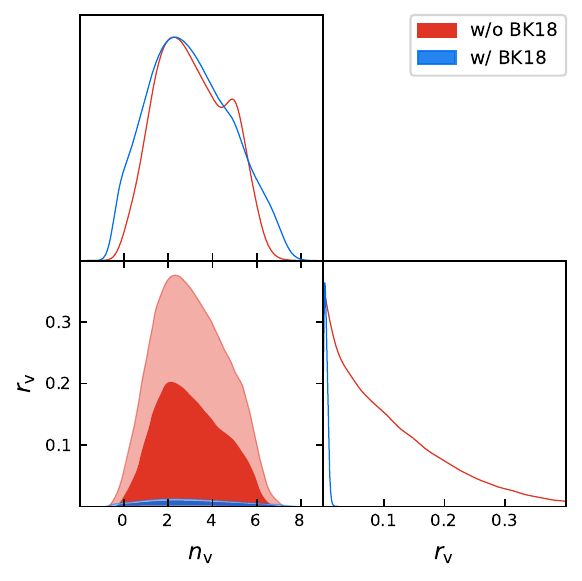}
    \caption{Constraints on the same parameters as Figure~\ref{fig:const_iso}, but for the $\noct$ mode.}
    \label{fig:const_oct}
\end{figure}

\begin{table}[t]
\centering
\renewcommand{\arraystretch}{1.25}
\begin{tabular}{c|c|c|c}
\hline\hline
Initial condition & Parameter & w/o BK18 & w/ BK18 \\
\hline
$\nvi$
  & $n_{\mathrm v}$ & $1.4^{+1.1}_{-1.1}$            & $0.8^{+1.9}_{-1.4}$            \\
  & $r_{\mathrm v}$ & $\lesssim 2.79 \times 10^{-3}$ & $\lesssim 1.55 \times 10^{-4}$ \\
  \hline
$\noct$
  & $n_{\mathrm v}$ & $3.1^{+2.9}_{-2.7}$            & $3.0^{+3.5}_{-3.3}$            \\
  & $r_{\mathrm v}$ & $\lesssim 0.307$               & $\lesssim 1.04 \times 10^{-2}$ \\
\hline\hline
\end{tabular}
\caption{Constraints on the primordial vector-mode parameters for the $\nvi$ and $\noct$ modes, with and without BK18 data.}
\label{tab:constraints}
\end{table}

Figures~\ref{fig:const_iso} and \ref{fig:const_oct} present the marginalized posterior contours in $n_{\mathrm{v}}$-$r_{\mathrm{v}}$ plane for the $\nvi$ and $\noct$ modes, respectively.
The corresponding results are summarized in Table~\ref{tab:constraints}. 
Constraints on the full set, including cosmological parameters, are presented in Appendix~\ref{sec:full_constraints}.
To isolate the impact of BK18, we performed the analysis with and without BK18 data. 
For both cases of the $\nvi$ and $\noct$ modes, the inclusion of BK18 $BB$ data significantly tightens the constraint on $r_{\mathrm{v}}$. 
We also find that the inclusion of BK18 shifts the preferred value of $n_{\mathrm{v}}$ toward smaller values. 
For blue spectra with $n_{\mathrm{v}}\gtrsim1$, increasing $n_{\mathrm{v}}$ enhances the vector contribution to the $BB$ spectrum at higher multipoles, particularly around $\ell\sim200$ where BK18 has high sensitivity. 
BK18, therefore, provides the dominant constraining power in this regime and strongly limits large values of $n_{\mathrm{v}}$.
Compared with the upper bound on $r_{\mathrm{v}}$ for the $\nvi$ mode, $r_{\mathrm{v}}\lesssim9.45\times10^{-3}$, reported in Ref.~\cite{2012PhRvD..85d3009I}, our constraint is tighter by about two orders of magnitude.

We evaluate the effect of changing the pivot scale from $k_{0}=0.01\, {\rm Mpc}^{-1}$ to $k_{0}=0.05\, {\rm Mpc}^{-1}$ on the upper bounds on $r_{\mathrm{v}}$.
We define the derived parameter by the following equation for the conversion as $r_{\rm v,\,0.05}=r_{\rm v}\times5^{n_{\rm v}-1}$.
Using the posterior samples, we obtain $r_{\rm v,\,0.05}\lesssim7.8\times10^{-5}$ and $r_{\rm v,\,0.05}\lesssim1.7$ for the $\nvi$ and $\noct$ modes, respectively.
As expected by the equation for the conversion, the effect on the derived parameter $r_{\rm v,\,0.05}$ gets larger for the larger $n_{\rm v}$.
Owing to the difference between the best-fit values of $n_{\rm v}$, the upper bounds on the derived parameter $r_{\rm v,\,0.05}$ get smaller and larger for the $\nvi$ and $\noct$ modes, respectively.

The constraints also exhibit a nontrivial dependence on the choice of initial conditions for the primordial vector mode. 
In particular, the preferred value of $n_{\mathrm{v}}$ for the $\noct$ mode differs from that for the $\nvi$ mode. 
Since, as shown in Fig.~\ref{fig:ISO_OCT_Cl}, the $BB$ spectrum sourced by the $\noct$ mode peaks around $\ell\approx100$, and its tilt for $n_{\mathrm{v}}=3.0$ is similar to that of the BK18 data, the best-fit value shows $n_{\mathrm{v}}=3.0$ for the $\noct$ mode.
As for $r_{\mathrm{v}}$, the upper bounds for the $\noct$ mode is less stringent than those for the $\nvi$ mode because for a fixed value of $r_{\mathrm{v}}$, the $\noct$ mode generates a smaller $BB$ spectrum than the $\nvi$ mode.
Note that, when we change the pivot scale for the $\nvi$ and $\noct$ modes to $k_0=0.05\,{\rm Mpc}^{-1}$, the corresponding multipoles shift to larger values. 
As a result, the preferred value of $n_{\mathrm{v}}$ increases only for the $\noct$ mode.
Consequently, the upper bound on $r_{\mathrm{v}}$ becomes significantly weaker for the $\noct$ mode, whereas we cannot observe a significant change for the $\nvi$ mode.

We also considered the tensor contributions to the total power spectrum, and the corresponding results are presented in Appendix~\ref{sec:vec_with_tensor}.
In that case, the upper bounds on $r_{\mathrm{v}}$ become slightly tighter than those in the analysis without tensor mode, yielding $r_{\mathrm{v}}\lesssim1.25\times10^{-4}$ and $r_{\mathrm{v}}\lesssim8.44\times10^{-3}$ for the $\nvi$ and $\noct$ modes, respectively.

\section{Observational implications}
\label{sec:Implications}

So far, we have focused on updating the constraints on the vector mode arising from the free-streaming neutrinos.
Here, we discuss the observational implications from such vector modes: generation of cosmic magnetic fields (\ref{sec:Mag}) and observed parity-violating signature in CMB measurements (\ref{sec:EB}).

\subsection{The strength of the induced magnetic fields}
\label{sec:Mag}

In the primordial plasma before the recombination epoch, the regular vector mode sources the relative vorticity between baryons and photons, and hence we expect the inevitable generation of magnetic fields through this mechanism. 
This section is devoted to investigating the possible amplitude of magnetic fields~\cite{1970MNRAS.147..279H,2006Sci...311..827I}.

The leading-order evolution equation for magnetic fields generated through Thomson scattering is given by~\cite{2012PhRvD..85d3009I}
\begin{align}
    \frac{\mathrm{d} (a^2B^i)}{\mathrm{d}t}
    =\frac{4\sigma_{\rm T}\rho_{\gamma}a}{3e}\epsilon^{ijk}
    \left(v_{\gamma j,k}-v_{\mathrm{b} j,k}\right) ,
    \label{eq: B evolv}
\end{align}
where the quantities $\epsilon^{ijk}$, $\sigma_{\rm T}$, $e$, $t$, $B^i$, $\rho_\gamma$, $v_{\gamma j}$, and $v_{\mathrm{b} j}$ are the Levi-Civita tensor, the Thomson scattering cross section, the elementary electric charge, the cosmic time, the magnetic-field vector, the photon energy density, the bulk velocity fields of photons, and that of baryons, respectively. 
A comma ${}_{,k}$ denotes the spatial derivative with respect to the spatial coordinate $x^{k}$.

Here, we use the tight coupling approximation of $v_{\gamma}-v_{\rm b}$: 
\begin{align}
    \label{eq:tight_coupling}
    v_{\gamma}-v_{\rm b}=\frac{R\mathcal{H}}{(1+R)\dot{\tau}}v_{\gamma}-\frac{4}{15}\frac{R}{1+R}\left(\frac{k}{\dot{\tau}}\right)^2(v_{\gamma}+\sigma),
\end{align}
where the quantities $\dot\tau=an_e\sigma_{\rm T}$ and $R=4\rho_{\rm b}/3\rho_{\gamma}$ are the differential optical depth and baryon-to-photon energy density ratio, respectively. 
For the $\nvi$ mode, the photon velocity remains constant at a leading order, whereas for the $\noct$ mode it is proportional to $\eta^3$.
Thus, the photon velocity term contributes to magnetic-field generation for the $\nvi$ mode, whereas its contribution is suppressed for the $\noct$ mode because $v_\gamma$ appears only at higher order in $\eta$.

From Eq.~\eqref{eq: B evolv} and \eqref{eq:tight_coupling}, we estimate the generated magnetic fields by~\cite{2025JCAP...10..112K}
\begin{align}
    a^2B\simeq \frac{\sigma_{\mathrm{T}}a^2 \rho_{\mathrm{b}}}{e}\left(\frac{k}{\dot\tau}\right) \sigma_0,
    \label{eq:solution_mag}
\end{align}
and
\begin{align}
    a^2B\simeq\frac{16\sigma_{\rm T}}{45e}\frac{a^2}{\mathcal{H}}\rho_{\rm b}k\left(\frac{k}{\dot{\tau}}\right)^2\sigma_0 ,
\end{align}
for the $\nvi$ and $\noct$ modes, respectively.
The quantity $\sigma_{0}$ denotes the initial value of the vector mode.
Here, we exploited the tight-coupling approximation valid in the deeply radiation-dominated universe (see Eq.~(32) in Ref.\cite{2012PhRvD..85d3009I} for details).

By approximating the initial amplitude of the vector modes as $\sigma_0\equiv\sqrt{\mathcal{P}_{\sigma}(k)}$ and adopting the best-fit value of $n_{\mathrm v}$ and the upper bounds on $r_{\mathrm v}$ (see Table~\ref{tab:constraints}), we estimate the amplitude of the resultant magnetic fields as
\begin{align}
    &a^2B\notag\\
    \simeq& 3.0\times 10^{-31} \, \mathrm{G} \notag\\
    \times&\left(\frac{k}{0.01\,\mathrm{Mpc}^{-1}}\right)^{\frac{n_{\rm v}+1}{2}} \biggl(\frac{a}{10^{-4}}\biggr) \left(\frac{r_{\mathrm{v}}}{1.55\times 10^{-4}}\right)^{1/2} ,
\end{align}
and 
\begin{align}
    &a^2B\notag\\
    \simeq &4.0\times 10^{-35} \, \mathrm{G} \notag\\
    \times&\left(\frac{k}{0.01\,\mathrm{Mpc}^{-1}}\right)^{\frac{n_{\rm v}+5}{2}} \biggl(\frac{a}{10^{-4}}\biggr)^{4} \biggl(\frac{r_{\mathrm{v}}}{1.04\times10^{-2}}\biggr)^{1/2} ,
\end{align}
for the $\nvi$ and $\noct$ modes, respectively.
Consequently, around the cosmological recombination epoch ($a\simeq10^{-3}$), we find upper bounds on the physical field strength of $B\simeq 10^{-23}\, \mathrm{G}$ and $B\simeq 10^{-21}\, \mathrm{G}$ at $k=1\,{\rm Mpc}^{-1}$ for the $\nvi$ and $\noct$ modes, respectively.
The former is about two orders of magnitude tighter than the $\mathcal{O}(10^{-21})\,\mathrm{G}$ bound reported in Ref.~\cite{2012PhRvD..85d3009I} for $\nvi$ mode due to a tighter constraint on $r_{\rm V}$, while the latter is comparable.
Such a field amplitude is far too small to act as a viable seed for the magnetic fields observed in galaxies and galaxy clusters.

\subsection{$EB$ power spectrum}
\label{sec:EB}

In this section, we consider the helical primordial vector modes to assess whether the observed $EB$ power spectrum~\cite{2020A&A...643A..42P,2023PhRvL.131l1001E} can be explained by a primordial origin, particularly vector perturbations.
We define the power spectra for the vector mode of each helicity state by
\begin{align}
\langle\sigma^{(\pm)}(\bm{k})\sigma^{(\pm)*}(\bm{k}')\rangle& = (2\pi)^3\delta_{\rm D}(\bm{k}-\bm{k}')\frac{2\pi^2}{k^3}\mathcal{P}^{(\pm)}_{\sigma}(k) .
\end{align}
Assuming fully helical primordial vector modes, we model the power spectra by
\begin{align}
\mathcal{P}^{(+)}_{\sigma}(k) & = A_{\sigma}\left(\frac{k}{k_0}\right)^{n_{\rm v}-1}, \\
\mathcal{P}^{(-)}_{\sigma}(k) & = 0 .
\end{align}

The $EB$ power spectrum sourced by the helical vector modes is given by
\begin{align}
    C_{\ell}^{EB}=\int d(\ln{k}) [\mathcal{P}_{\sigma}^{(+)}(k)-\mathcal{P}_{\sigma}^{(-)}(k)]\Delta_{\ell}^{E}(k)\Delta_{\ell}^{B}(k),
\end{align}
where $\Delta^{E/B}_{\ell}$ is the transfer function of $E$-mode and $B$-mode polarizations of the CMB photons, respectively.
Note that the parity-even observable is proportional to $\mathcal{P}^{(+)}_{\sigma}+\mathcal{P}^{(-)}_{\sigma}$, while the parity-odd one is proportional to $\mathcal{P}^{(+)}_{\sigma}-\mathcal{P}^{(-)}_{\sigma}$.
If we interpret the constraints on the vector-modes parameters we obtained in Sec.~\ref{sec:Constraints} as those on helical modes, we can compute the $EB$ power spectrum with \texttt{CAMB}.
We present the $EB$ power spectra from the helical vector modes and those measured by \textit{Planck}.
We calculate the power spectrum for both the $\nvi$ and $\noct$ modes.
We have taken $r_{\mathrm{v}}$ as the value of the upper bounds found in Sec.~\ref{sec:Constraints}, which makes the amplitude of the $EB$ spectrum maximum.

\begin{figure}[t]
  \centering
  \includegraphics[width=0.99\linewidth]{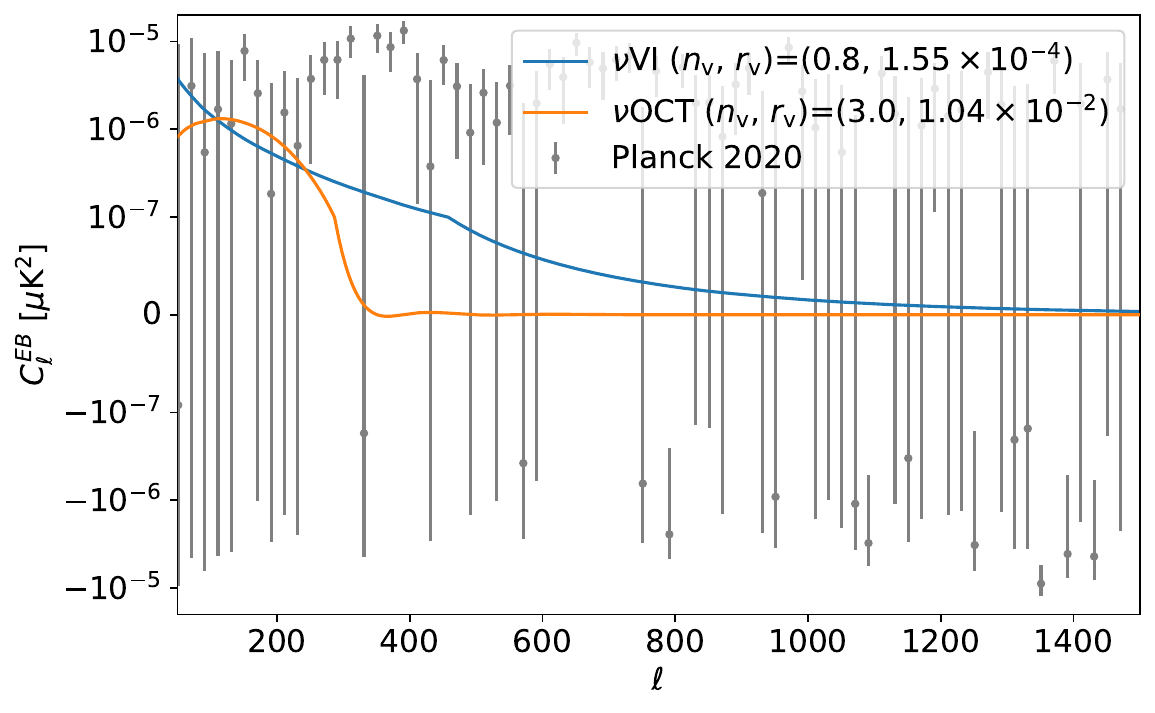}
  \caption{$EB$ power spectra from fully helical primordial vector modes and the observed spectrum reported by \textit{Planck}~\cite{2020A&A...643A..42P,2023PhRvL.131l1001E}. 
  Blue and orange curves show the $EB$ spectra from helical vector modes for the $\nvi$ (blue curve) and $\noct$ (orange curve) modes, respectively. 
  For both modes, $n_{\mathrm{v}}$ and $r_{\mathrm{v}}$ are fixed to the best-fit values and the upper bounds, respectively. 
  Note that the theoretical spectra sourced by the fully helical vector modes are multiplied by $10^{2}$ for visualization purposes.}
  \label{fig:eb_ISO_OCT}
\end{figure}

\begin{figure}
    \includegraphics[width=0.99\linewidth]{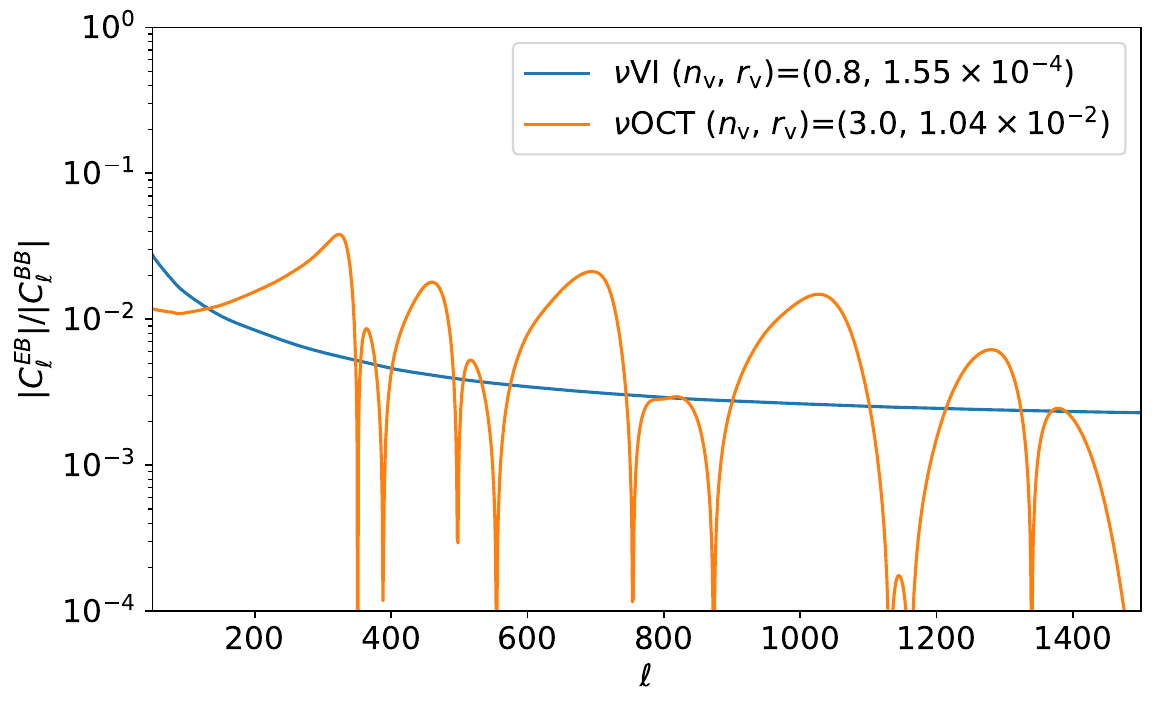}
    \caption{Ratio of the $EB$ to the $BB$ power spectra. 
    As indicated in the figure, $n_{\mathrm{v}}$ and $r_{\mathrm{v}}$ are fixed at the best-fit values and the upper bounds, respectively.}
    \label{fig:eb_bb_ratio}
\end{figure}

Figure~\ref{fig:eb_ISO_OCT} shows the theoretical $EB$ power spectra from the $\nvi$ and $\noct$ modes, together with the spectrum measured by \textit{Planck}~\cite{2023PhRvL.131l1001E,2020A&A...643A..42P}. 
Here, $n_{\mathrm{v}}$ and $r_{\mathrm{v}}$ are fixed to the best-fit value and the upper bounds, respectively.
Note that for visualization purposes, the theoretical spectra sourced by the fully helical vector modes are multiplied by $10^{2}$.
As seen in Fig.~\ref{fig:eb_ISO_OCT}, even when $n_{\mathrm{v}}$ and $r_{\mathrm{v}}$ are set to their best-fit values and the upper bounds obtained for each mode, respectively, the $EB$ power spectra sourced by the helical vector modes are inconsistent in both amplitudes and shapes with the observed $EB$ spectrum.
With the MCMC results, the vector mode typically produces a sizable $EB$ signal on larger angular scales than those at which the observed non-zero $EB$ spectrum is reported. 
Also, more importantly, the amplitude of the $EB$ spectrum induced by the vector mode is much smaller than the observed one.
This smallness of the produced $EB$ spectrum can be understood by looking at the ratio of the $EB$ spectra to the $BB$ spectra for the best-fit value of $n_{\mathrm{v}}$ and the upper bounds of $r_{\mathrm{v}}$ depicted in Fig.~\ref{fig:eb_bb_ratio}.
For both the $\nvi$ and $\noct$ modes, the $BB$ spectra are larger in amplitude than the $EB$ spectra over almost the entire multipole range.
As discussed in Sec.~\ref{sec:Constraints}, the constraints on the vector modes are mainly driven by the $BB$ power spectrum.
This in turn implies that helical primordial vector modes are generally insufficient to explain the measured $EB$ spectrum while remaining consistent with the observed $BB$ power spectrum.
We leave the discussion of the vector-mode case of the log-normal initial power spectrum to our future work.

\section{Summary}
\label{sec:Summary}

In this paper, we have revisited observational constraints on primordial vector modes induced by the neutrino-sector initial conditions.
Using current cosmological data, we constrained the amplitude and spectral index of the vector mode and examined the implications of these bounds for primordial magnetic-field generation prior to recombination.
In particular, we obtained the upper bounds $r_{\mathrm{v}}\lesssim1.55\times10^{-4}$ and $r_{\mathrm{v}}\lesssim1.04\times10^{-2}$ for the $\nvi$ and $\noct$ modes by including BK18 $B$-mode polarization data, which is more stringent than the previous result of Ref.~\cite{2012PhRvD..85d3009I}.
These constraints imply that the magnetic fields inevitably generated through the baryon-photon relative velocity reach only $B\simeq10^{-23}\,{\rm G}$ and $B\simeq10^{-21}\,{\rm G}$ around the recombination epoch for the $\nvi$ and $\noct$ modes, respectively.
Such amplitudes are too weak to account for the seed fields required for the magnetic fields observed today.

We also extended the analysis to the helical case and calculated the $EB$ power spectra sourced by fully helical primordial vector modes.
This enables a direct test of whether such vector perturbations can account for the measured parity-odd polarization signal. 
We found that, although helical vector modes generate a non-vanishing $EB$ correlation, their required amplitude is tightly constrained mainly by the $B$-mode polarization measurements, significantly limiting their contribution to the $EB$ signal.

These results strengthen the observational constraints on primordial vector modes and on the magnetogenesis scenarios associated with them. 
More broadly, our analysis highlights that parity-even observables constrain the overall vector amplitude, while parity-odd observables such as $EB$ spectra provide an additional and complementary probe of primordial helicity. 
This demonstrates the importance of combining conventional cosmological constraints with parity-odd observables in searching for new physics in the primordial universe.

\section*{Note added}

While this work was being completed, we learned that A.~R.~Khalife and C.~Pitrou were working on a related study, where the constraints on the vector-mode initial conditions are investigated using a comprehensive MCMC analysis including recent CMB polarization data~\cite{2026arXiv260508907R}. The present work has a different focus. 
We obtain the updated constraint on the magnetic fields induced from the primordial vector modes. In addition, by performing a complementary analysis of the $EB$ polarization spectrum, we discuss the possibility of parity violation in the vector sector, concentrating on the $\nvi$ and $\noct$ modes.

\begin{acknowledgments}
We thank A.~R.~Khalife and C.~Pitrou for helpful discussions and for kindly communicating the status of their ongoing work. This work was supported in part by the Japan Society for the Promotion of Science (JSPS) KAKENHI Grant Numbers JP24K17043 (SS), JP23H00108 (SY), JP24K00627 (SY), and 21H04467 (KI), 24K00625 (KI), and JST FOREST Program JPMJFR20352935 (KI). SS also acknowledges support from DAIKO FOUNDATION and Hirose Foundation.
\end{acknowledgments}

\appendix

\section{Full constraints on the cosmological parameters and vector mode}
\label{sec:full_constraints}

In this appendix, we show the full constraints on the standard $\Lambda$CDM parameters and the primordial vector-mode parameters, $n_{\mathrm{v}}$ and $r_{\mathrm{v}}$. 

\begin{figure*}[t]
    \centering
    \includegraphics[width=0.95\linewidth]{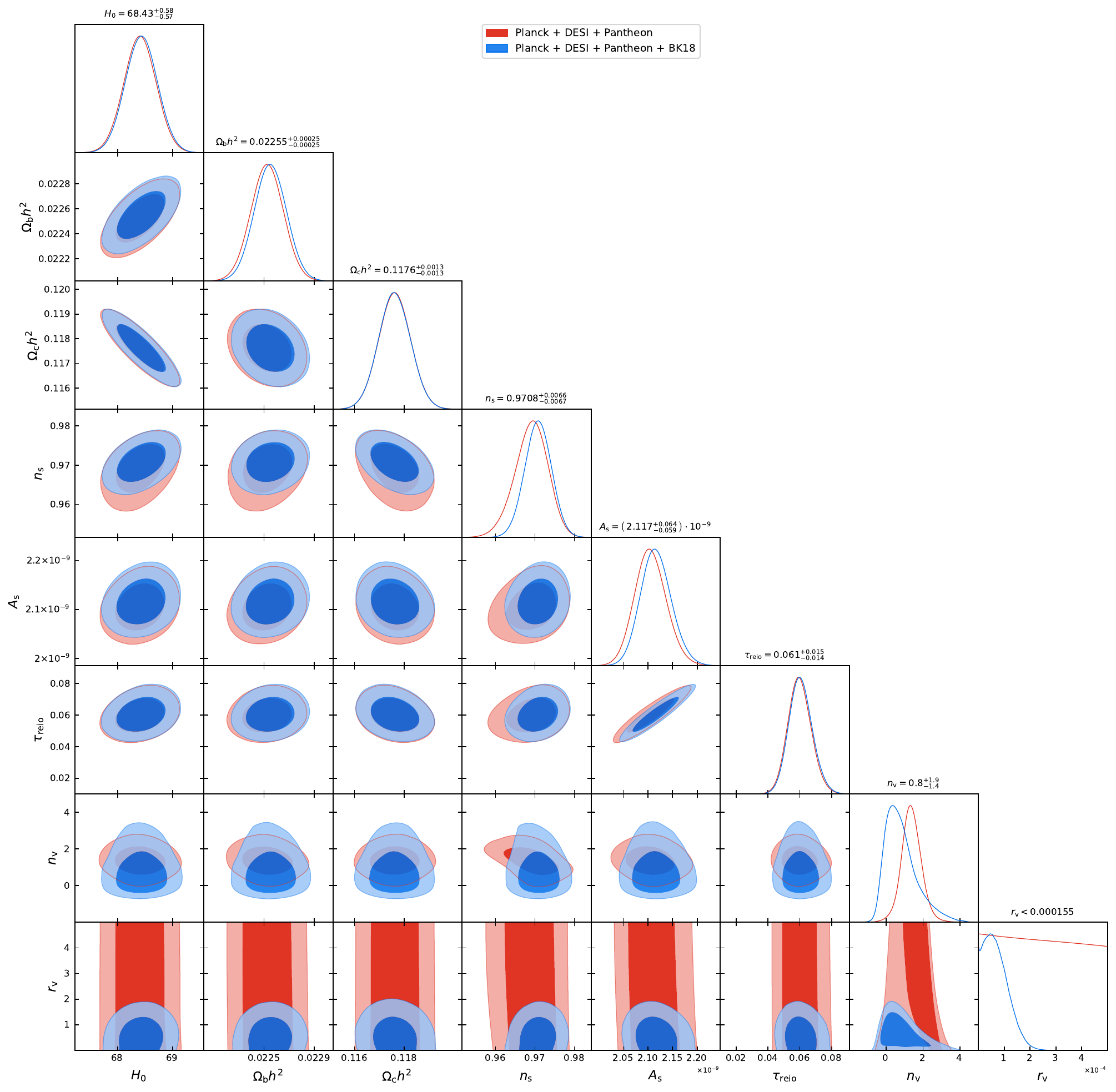}
    \caption{Constraints on the standard $\Lambda$CDM model parameters and the $\nvi$-mode parameters, $n_{\mathrm{v}}$ and $r_{\mathrm{v}}$. 
    The red and blue contours represent the results without and with BK18 data, respectively. 
    The title of each panel shows the best-fit value and the 95\% confidence interval for the case with BK18.
    Since the posterior distribution of $n_{\rm v}$ is non-Gaussian, and the MCMC chains accumulate in its right tail, the contours in the $n_{\mathrm{v}}$-$r_{\mathrm{v}}$ plane can misleadingly suggest that $r_{\mathrm{v}}$ is inconsistent with zero.}
    \label{fig:constraints_ISO_full}
\end{figure*}

\begin{figure*}[t]
    \centering
    \includegraphics[width=0.95\linewidth]{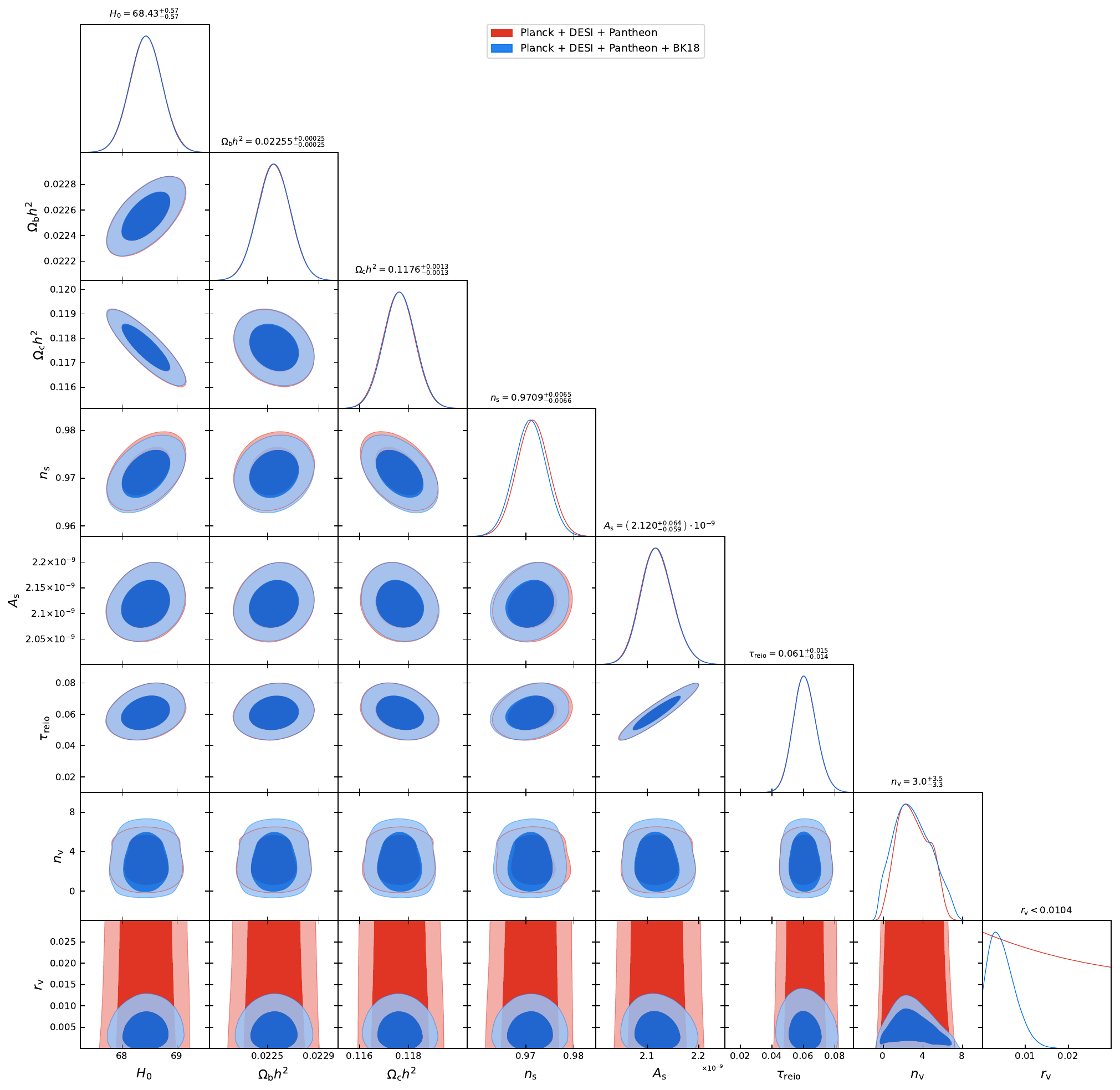}
    \caption{Same as Figure~\ref{fig:constraints_ISO_full}, but for the $\noct$-mode.}
    \label{fig:constraints_OCT_full}
\end{figure*}

Figures~\ref{fig:constraints_ISO_full} and \ref{fig:constraints_OCT_full} present the corresponding triangle plots from the full parameter analysis for the $\nvi$ and $\noct$ modes, respectively.
We note the best-fit value and the 95\% confidence interval for the corresponding parameter at the top of each panel.
Overall, we confirm that the standard cosmological parameters are not significantly affected by the inclusion of the primordial vector modes.
Interestingly, while the $\nvi$ mode slightly affects the constraint on $n_{\rm s}$, the $\noct$ does not.
This small apparent shift in $n_{\rm s}$ is consistent within 2-$\sigma$ errors, though.
This is due to the bluer slope of the $BB$ spectrum at low multipoles induced by the $\noct$ mode, which suppresses its impact on the $TT$ spectrum at low multipoles. 
Therefore, compared to $\nvi$, the $\noct$ mode is more disentangled from the scalar-sector parameters.
In the $n_{\mathrm{v}}$-$r_{\mathrm{v}}$ plane, since the posterior distribution of $n_{\rm v}$ is non-Gaussian, and the MCMC chains accumulate in its right tail, $r_{\mathrm{v}}$ seems misleadingly to be inconsistent with zero.

\section{Constraints on the vector mode in the presence of tensor modes}
\label{sec:vec_with_tensor}

In this appendix, we show the constraints on the $\nvi$ and $\noct$ modes when primordial tensor modes are also included in the analysis. 
The parameterization of the tensor mode initial power spectrum we use in the analysis follows: 
\begin{align}
    \mathcal{P}_{\rm t}(k)=A_{\rm t}\left(\frac{k}{k_{\rm t0}}\right)^{n_{\rm t}},
\end{align}
where the pivot scale $k_{\rm t0}=0.05\mathrm{Mpc}^{-1}$.
We define the tensor-to-scalar ratio as
\begin{align}
    r_{0.05}=\frac{A_{\rm t}}{A_{\rm s}},
\end{align}
where $n_{\mathrm{t}}$ and $r_{0.05}$ satisfy the consistency relation, $n_{\mathrm{t}}=-r_{0.05}/8$.

\begin{figure}
    \centering
    \includegraphics[width=0.99\linewidth]{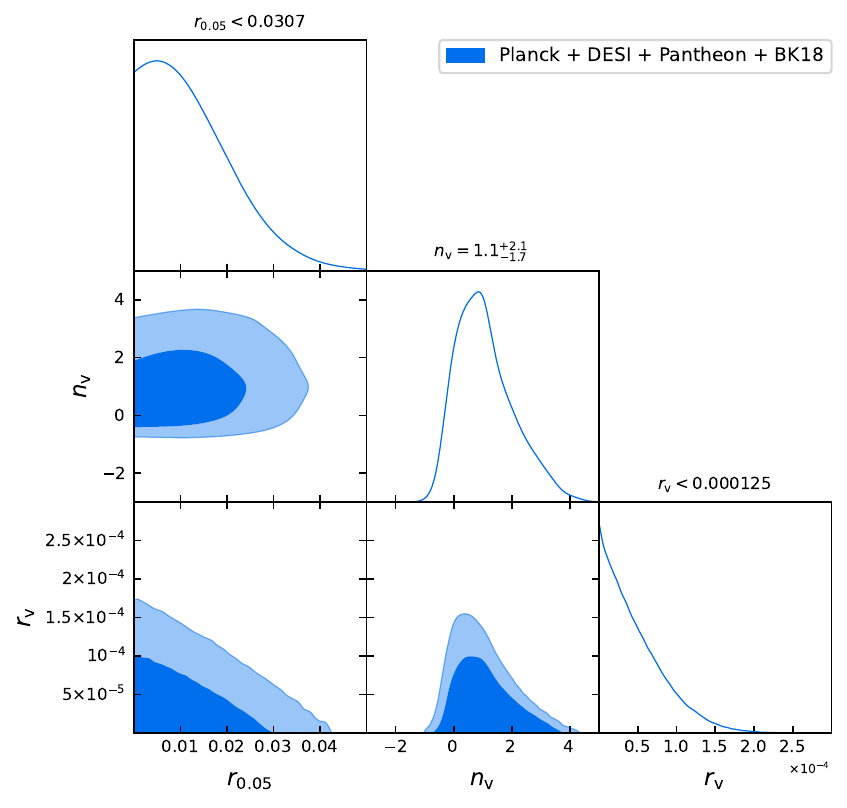}
    \caption{The constraints on the parameters of the vector and tensor mode, $n_{\mathrm{v}}$, $r_{\mathrm{v}}$, and $r_{0.05}$ at 68\% and 95\% confidence levels. The title of each panel gives the best-fit value and the 95\% confidence interval for the corresponding parameter.}
    \label{fig:constraints_vec_tensor_ISO}
\end{figure}

\begin{figure}
    \centering
    \includegraphics[width=0.99\linewidth]{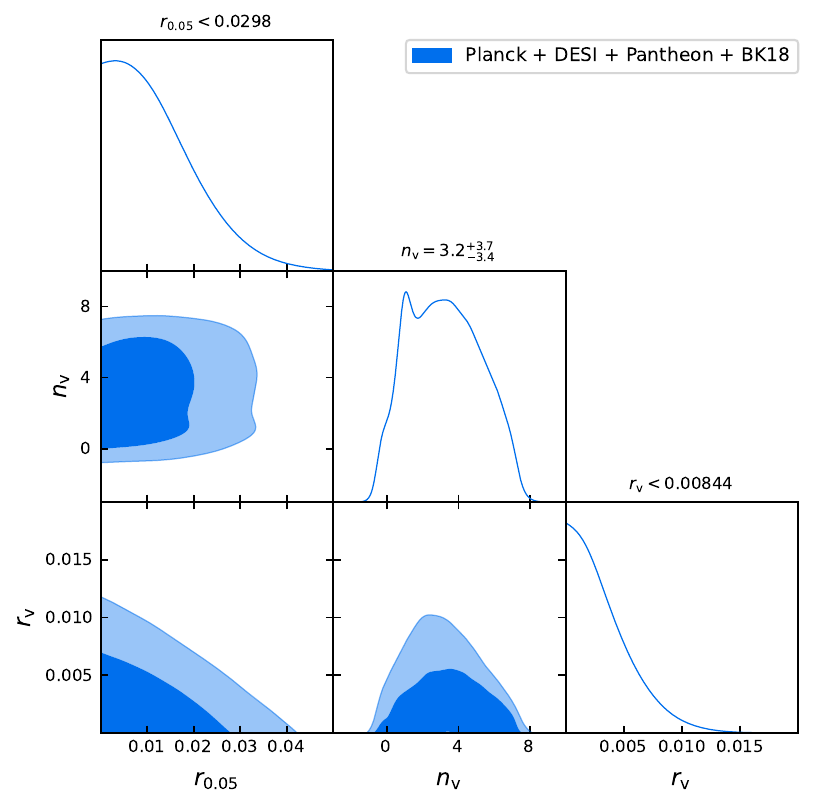}
    \caption{Same as Figure~\ref{fig:constraints_vec_tensor_ISO}, but for $\noct$ mode.}
    \label{fig:constraints_vec_tensor_OCT}
\end{figure}

Figures~\ref{fig:constraints_vec_tensor_ISO} and \ref{fig:constraints_vec_tensor_OCT} present the corresponding 
triangle plots for the $\nvi$ and $\noct$ modes, respectively.
We find that the allowed region of $r_{\mathrm{v}}$ becomes tighter in this extended analysis, reflecting the fact that the data constrain the combined vector-plus-tensor contribution to the $BB$ spectrum. 
The posteriors show clear correlations between $r_{\mathrm{v}}$ and $r_{0.05}$, approximately given by
\begin{align}
  r_{\mathrm{v}} + 0.004 r_{0.05} \lesssim 1.75\times10^{-4},
\end{align}
and
\begin{align}
     r_{\mathrm{v}} + 0.3 r_{0.05} \lesssim 0.0125,
\end{align}
for the $\nvi$ and $\noct$ modes, respectively.
These relations imply that the data constrain not the vector and tensor amplitudes independently, but rather a particular combination of them. 
As a result, the allowed parameter space is restricted along a narrow degeneracy direction, leading to a tighter upper bound on $r_{\mathrm{v}}$ than in the analysis without tensor modes.

\bibliography{ref}

\end{document}